\documentclass[preprint,showpacs,preprintnumbers,amsmath,amssymb,aip]{revtex4-1}

\usepackage{graphicx}
\usepackage{dcolumn}
\usepackage{bm}
\usepackage{psfrag}

\begin{document}

\preprint{APS/123-QED}

\title{Bounds for rotating convection at infinite Prandtl number from
semidefinite programs}

\author{A. Tilgner}

\affiliation{Institute of Astrophysics and Geophysics, University of G\"ottingen,
Friedrich-Hund-Platz 1, 37077 G\"ottingen, Germany }

\date{\today}

\begin{abstract}
Bounds for the poloidal and toroidal kinetic energies and the heat transport are
computed numerically for rotating convection at infinite Prandtl number with
both no slip and stress free boundaries. The constraints invoked in this
computation are linear or quadratic in the problem variables and lead to the
formulation of a semidefinite program. The bounds behave as a function of
Rayleigh number at fixed Taylor number qualitatively in the same way as the
quantities being bounded. The bounds are zero for Rayleigh numbers smaller
than the critical Rayleigh number for the onset of convection, they increase
rapidly with Rayleigh number for Rayleigh numbers just above onset, and increase
more slowly at large Rayleigh numbers. If the dependencies on Rayleigh number
are approximated by power laws, one obtains larger exponents from bounds on the
Nusselt number for Rayleigh numbers just above onset than from the actual Nusselt
number dependence known for large but finite Prandtl number. The wavelength of
the linearly unstable mode at the onset of convection appears as a relevant
length scale in the bounds.
\end{abstract}

\pacs{47.27.te, 47.27.-i, 44.25.+f}
\maketitle

\section{Introduction}

The optimum theory of turbulence attempts to replace the exact computation of
time averaged global quantities, such as the heat transfer across a convecting
layer, by a simpler calculation at the price of obtaining only upper bounds to
the quantities of interest. The method most frequently used for this purpose is
the background field method \cite{Hopf41,Doerin92,Doerin96,Fantuz22}, which is
more convenient to use but which ultimately produces the same results as another
method based on refs. \onlinecite{Malkus56,Howard63,Busse69}, at least for the
convection problem \cite{Kerswe01}. Both classes of methods do not make use of the full
Navier-Stokes equation but use relations derived from integrals of this
equation, such as the energy budget. As a consequence, a major disadvantage of
these methods applied to convection flows is that they are insensitive to the
presence of the Coriolis term in the momentum equation. Rotational effects
appear in the bounds derived from straightforward optimum theory of convection
only for reduced sets of equations \cite{Grooms14,Pachev20}, the simplest representative
of which is the model of the infinite Prandtl number.

Rotating convection is a problem with many applications in geophysics and
astrophysics. In this context, convection at large Prandtl numbers in particular also plays a
role. Solutal convection in the oceans or in planetary cores \cite{Jones15} is characterized by
Prandtl numbers around $10^3$ or $10^4$. It is not obvious if, or under which
conditions, rotating convection with Prandtl numbers this large behaves the same as
convection in a fluid of infinite Prandtl number. However, these applications
motivate the study of convection at infinite Prandtl number as a simple limiting
case. In addition, it was recently shown \cite{Tilgne22} how bounds derived for
infinite Prandtl number convection can be used as a starting point for the
computation of bounds for arbitrary Prandtl number. This calculation can be
summarized as a background field method with a background field for velocity
which is not fixed in advance and time independent, but which instead is the
solution of the momentum equation for infinite Prandtl number. The bounds
derived in this manner for general Prandtl number cannot be lower than the
bounds for infinite Prandtl number. It is therefore interesting to explore
numerically which are the best possible bounds derivable at infinite Prandtl number
from the constraints used in optimum theory of turbulence as we presently know it.

The known bounds derived by analytical means \cite{Doerin01,Consta99b,Tilgne22}
for the heat transfer in rotating infinite Prandtl number convection 
are too loose to be satisfactory. While these bounds qualitatively
reproduce a reduction of the heat transport by rotation, they do not restrict
the advective heat transfer to zero for Rayleigh numbers below the critical
Rayleigh number computed from linear stability analysis.

A numerical computation of upper bounds for heat transfer in rotating convection
at infinite Prandtl number with stress free boundaries previously appeared in
ref. \onlinecite{Vitano03}. The parameter range covered in that study is too
small to allow us to really judge the performance of the bounds by comparing them with
experimental data. We also know from the non rotating problem that the optimum
theory for infinite Prandtl number behaves very differently for stress free and
no slip boundaries \cite{Plasti05,Ierley06}, and the boundary conditions
relevant for experiments are no slip rather than stress free boundaries.
Finally, ref. \onlinecite{Vitano03} did not compute bounds on energies.

There are thus several motivations to revisit the problem of determining
numerically the optimal bounds for rotating convection at infinite Prandtl
number. The computation presented here makes use of the same constraints as
previous work. These constraints only include linear
and quadratic terms. This problem can be cast in the form of a semidefinite
program (SDP). Numerical techniques of SDP have already been applied to
convection problems \cite{Tilgne17b,Tilgne19} and they promise to be more
generally applicable to bounding hydrodynamic quantities
\cite{Cherny14,Fantuz22}. The SDP implemented for rotating convection at infinite
Prandtl number is explained in the next section, and the results are
presented in section 3.

\section{The Semidefinite Program}
\label{section:SDP}

We consider a plane layer infinitely extended in the $x$ and $y$ directions and of
thickness $h$ with bounding planes perpendicular to
the $z$ axis in a frame of reference rotating with angular velocity $\Omega$
about the $z$ axis. 
The gravitational acceleration acts in the direction of negative $z$. The layer is
filled with fluid of density $\rho$, kinematic viscosity $\nu$, thermal
diffusivity $\kappa$, and thermal expansion coefficient $\alpha$. Top and bottom
boundaries are held at the fixed temperatures $T_\mathrm{top}$ and
$T_\mathrm{top} + \Delta T$, respectively. Within the Boussinesq approximation,
there are in general three control parameters: the Prandtl, Rayleigh and Taylor
numbers $\mathrm{Pr}$, $\mathrm{Ra}$ and $\mathrm{Ta}$ defined as
\begin{equation}
\mathrm{Pr}=\frac{\nu}{\kappa} ~~~,~~~ 
\mathrm{Ra}=\frac{g \alpha \Delta T h^3}{\kappa \nu} ~~~,~~~
\mathrm{Ta}=\frac{4 \Omega^2 h^4}{\nu^2}.
\end{equation}
Here, we immediately specialize to the limit of infinite $\mathrm{Pr}$ in which
case the equations of evolution may be written as follows in terms of the
non dimensional fields of velocity $\bm v(\bm r,t)$, pressure $p(\bm r,t)$ and
temperature deviation from the conductive temperature profile $\theta(\bm r,t)$
where $\theta(\bm r,t)=T(\bm r,t)-1+z$ with $T(\bm r,t)$ being the temperature:

\begin{eqnarray}
\sqrt{\mathrm{Ta}} \bm{\hat z} \times \bm v &=& 
-\nabla p + \mathrm{Ra} \theta \bm{\hat z} + \nabla^2 \bm v
\label{eq:NS} \\
\partial_t \theta + \bm v \cdot \nabla \theta -v_z &=& \nabla^2 \theta
\label{eq:Temp} \\
\nabla \cdot \bm v &=& 0
\label{eq:conti} 
\end{eqnarray}
$\bm{\hat z}$ denotes the unit vector in $z$ direction. The conditions at the boundaries 
$z=0$ and $z=1$ on the temperature imply that $\theta=0$ there.
Both stress free and no slip conditions will be investigated. At
stress free boundaries,
$\partial_z v_x = \partial_z v_y = v_z = 0$, whereas
$\bm v =0$ on no slip boundaries. Periodic boundary conditions are assumed in
the horizontal directions with arbitrary periodicity lengths.

We now decompose $\bm v$ into poloidal and toroidal scalars $\phi$
and $\psi$ such that
$\bm v = \nabla \times \nabla \times (\phi \bm{\hat z}) + \nabla \times (\psi \bm{\hat z})$
which automatically fulfills $\nabla \cdot \bm v = 0$. An additional term with a
mean flow is not necessary at infinite Prandtl number \cite{Tilgne22}. The $z$ component of
the curl and the $z$ component of the curl of the curl of eq. (\ref{eq:NS})
yield the equations of evolution for $\phi$ and $\psi$,

\begin{eqnarray}
\sqrt{\mathrm{Ta}} \partial_z \Delta_2 \psi
&=&
\nabla^2 \nabla^2 \Delta_2 \phi - \mathrm{Ra} \Delta_2 \theta 
\label{eq:phi}  \\
-\sqrt{\mathrm{Ta}} \partial_z \Delta_2 \phi
&=&
\nabla^2 \Delta_2 \psi
\label{eq:psi} 
\end{eqnarray}
with $\Delta_2 = \partial^2_x + \partial^2_y$. The
boundary conditions become
$\phi = \partial^2_z \phi = \partial_z \psi = 0$  for stress free boundaries and
$\phi = \partial_z \phi = \psi = 0$ for no slip boundaries.

The eqs. (\ref{eq:phi}) and (\ref{eq:psi}) determine $\phi$ and $\psi$, and hence
$\bm v$, in dependence of $\theta$. The equations of evolution
(\ref{eq:NS}-\ref{eq:conti}) therefore reduce to the single equation
(\ref{eq:Temp}) in which $\bm v$ ultimately is some function of $\theta$. Since
there is no closed expression for this function, we keep the variables $\phi$,
$\psi$ and $\bm v$ in subsequent formulae bearing in mind that they are nothing
but functions of $\theta$.

We next define averages over space of a function $f(\bm r)$ and over time of a
function $g(t)$ with the symbols
\begin{equation}
\overline{f(t)} = \lim_{\tau \rightarrow \infty} \frac{1}{\tau} \int_0^\tau f(t) dt
~~~,~~~
\langle g(\bm r) \rangle = \frac{1}{V} \int g(\bm r) dV
\end{equation}
where $V$ is the volume of a periodicity volume.
The product of eq. (\ref{eq:Temp}) with $\theta$ followed by a volume average
leads in this notation to
\begin{equation}
\partial_t \langle \frac{1}{2} \theta^2 \rangle = 
\langle v_z \theta \rangle - \langle |\nabla \theta|^2 \rangle
.
\label{eq:theta2}
\end{equation}

The quantities for which we will seek bounds in this paper are the time and
volume averaged heat transport $\langle \overline {v_z \theta} \rangle$,
the poloidal energy
\begin{equation}
E_\mathrm{pol} = 
\langle \frac{1}{2} \overline{|\nabla \times \nabla \times (\phi \bm{\hat z})|^2} \rangle
\label{eq:Epol}
\end{equation}
and the toroidal energy
\begin{equation}
E_\mathrm{tor} = 
\langle \frac{1}{2} \overline{|\nabla \times (\psi \bm{\hat z})|^2} \rangle
.
\label{eq:Etor}
\end{equation}
It only makes sense to ask for bounds on energy in convection with stress free
boundaries if one specifies a certain frame of reference because any solution to
the equations of evolution may be transformed to another solution with stress
free boundary conditions simply by changing to another frame of reference moving at
an arbitrarily large horizontal translation velocity. We adopt the same
convention as in Ref. \onlinecite{Tilgne22} and select the frame of reference in
which total momentum is zero.

The method to find these bounds is exactly the same as in Ref.
\onlinecite{Tilgne17b}. The only difference in the numerical implementation is
that in the non rotating problem of Ref. \onlinecite{Tilgne17b}, the toroidal
scalar $\psi$ was always zero, whereas here, both $\phi$ and $\psi$ have to be
computed in terms of $\theta$ taking into account a $\mathrm{Ta}$ different from
zero. The principle of the method is summarized here to make the paper self
contained.

We select test functions $\varphi_n(z)$, $n=1...N$ which depend on $z$ only and project onto them
the temperature equation (\ref{eq:Temp}):
\begin{equation}
\partial_t \langle \varphi_n \theta \rangle = 
\langle \varphi_n \partial_z (\theta \Delta_2 \phi) \rangle
+ \langle \varphi_n \nabla^2 \theta \rangle
.
\label{eq:projection}
\end{equation}
We now build a functional $G(\lambda_1 ... \lambda_N,\lambda_R,\theta)$ as a
linear combination of the right hand sides of the equations
(\ref{eq:projection}) and (\ref{eq:theta2}):
\begin{equation}
G(\lambda_1 ... \lambda_N,\lambda_R,\theta)=
\sum_{n=1}^N \lambda_n \left[ \langle \varphi_n \partial_z (\theta \Delta_2 \phi) \rangle
+ \langle \varphi_n \nabla^2 \theta \rangle \right]
+ \lambda_R \left[
\langle \theta \Delta_2 \phi \rangle + \langle |\nabla \theta|^2 \rangle
\right]
\label{eq:G}
.
\end{equation}
Let us call $Z$ the function for whose time average we want to find a bound. For example,
$Z = \langle v_z \theta \rangle$ if we want to compute a bound for the heat
transport. $Z$ is a quadratic functional of $\theta$ for the problems of
bounding heat transport, $E_\mathrm{pol}$ and $E_\mathrm{tor}$. Let us suppose
we know a number $\lambda_0$ and Lagrange multipliers $\lambda_1 ...
\lambda_N,\lambda_R$ such that all fields $\theta(\bm r)$ obeying the boundary
conditions for $\theta$ satisfy the following inequality:
\begin{equation}
-Z+\lambda_0+G(\lambda_1 ... \lambda_N,\lambda_R,\theta) \geq 0
.
\label{eq:p7Mitte}
\end{equation}
If we take the time average of this inequality keeping in mind that $G$ is a
linear combination of expressions equal to a time derivative, we find 
$\overline Z \leq \lambda_0$. 

The argument may equally well be recast in the form of a background method or
in terms of an auxillary function \cite{Fantuz22}. If we call
$\varphi(z)=\sum_{n=1}^N \lambda_n \varphi_n(z)$, we may write $G$ as 
\begin{equation}
G(\lambda_1 ... \lambda_N,\lambda_R,\theta)
=
\langle \varphi \partial_t \theta \rangle - 
\lambda_R \partial_t \langle \frac{1}{2} \theta^2 \rangle
=
- \frac{\lambda_R}{2} \frac{d}{dt} 
\langle \left( \theta - \frac{\varphi(z)}{\lambda_R} \right)^2 \rangle
\end{equation}
so that the background field $\tau(z)$ used within the background field method
applied to convection \cite{Doerin96} is identical with \cite{Tilgne17b}
$1-z+\varphi(z)/\lambda_R = \tau(z)$.

The best possible bound on $\overline Z$ is obtained by
minimizing $\lambda_0$ over the $\lambda_1 ... \lambda_N,\lambda_R$ subject to
the constraint (\ref{eq:p7Mitte}). This optimization problem turns into an SDP
after discretization of $\theta$. The computations presented here
discretized $\theta$ with $N$ Chebyshev polynomials in $z$ and a Fourier
decomposition in $x$ and $y$. Resolutions with $N$ up to 512 were used. The
package cvxopt provided the numerical solutions of the SDP. The details of 
the numerical code will not be described here as they are exactly the same as
in Ref. \onlinecite{Tilgne17b}. The technical points explained in this reference
are the symmetry of the problem about $z=1/2$, an automated search for
an active set of wavenumbers in the Fourier decomposition in the horizontal plane,
the exact formulation of the boundary conditions on $\theta$, a partial integration
of eq. (\ref{eq:G}) and a rescaling of $\lambda_0$ with powers of $\mathrm{Ra}$.

\begin{figure}
\includegraphics[width=8cm]{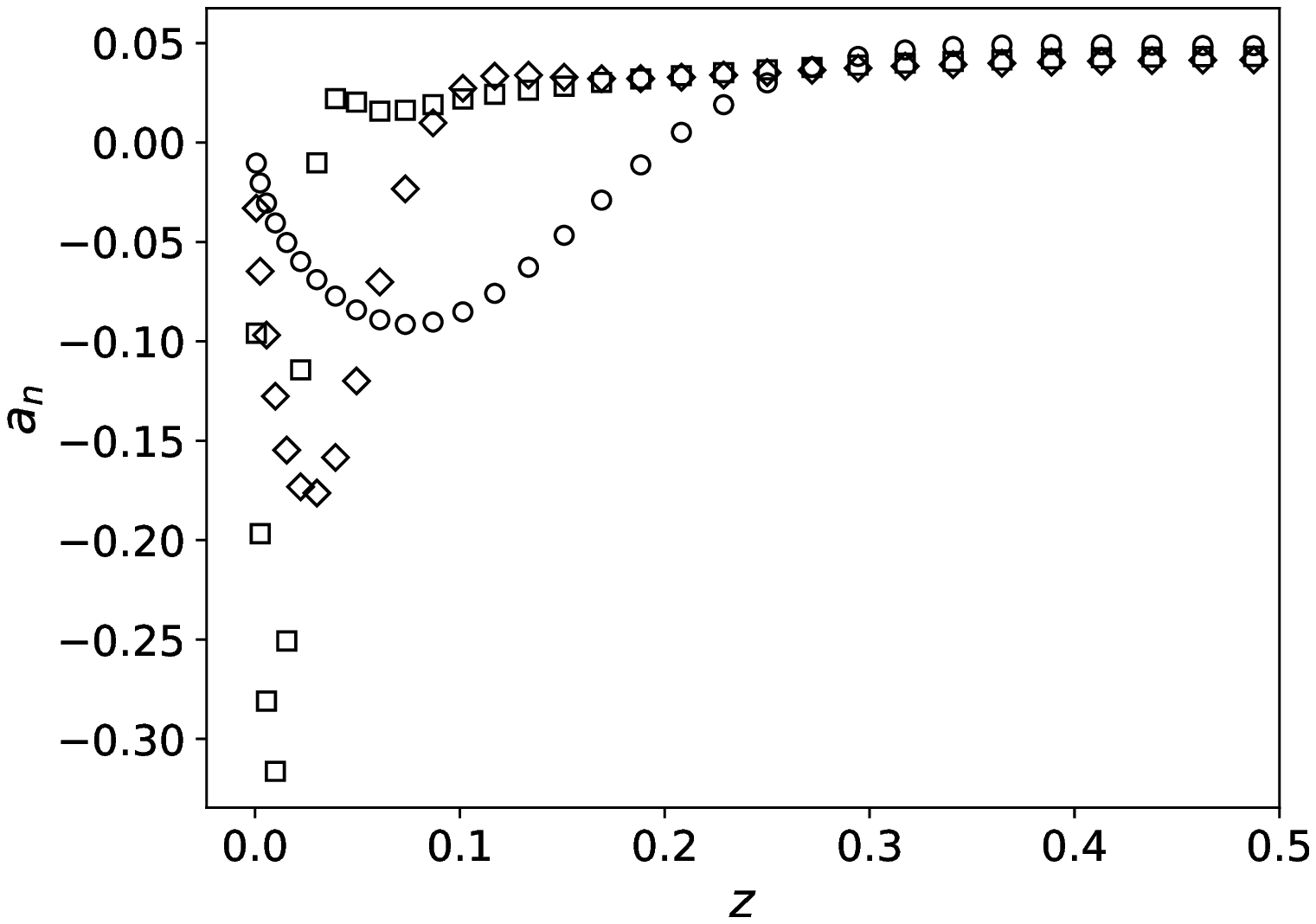}
\includegraphics[width=8cm]{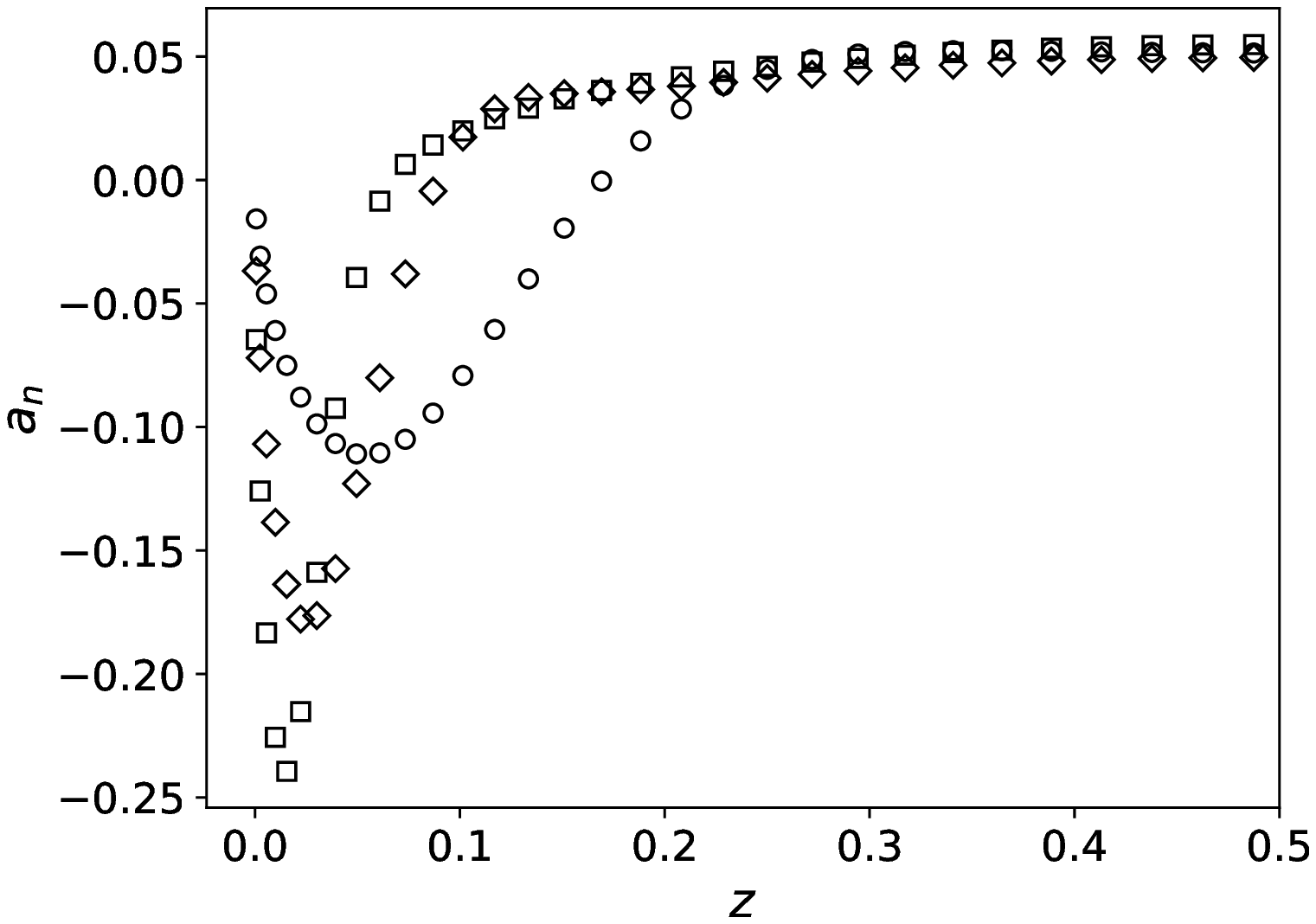}
\caption{
Coefficients $a_n$, plotted at locations $z_n$, for
$\mathrm{Ta}=10^4$ (circles), $10^6$ (diamonds) and $10^8$ (squares) and a 
resolution $N=64$. The boundary conditions are no slip (left panel) and stress 
free (right panel). The Rayleigh number is in all cases approximately 4 times
critical: $\mathrm{Ra}=34.92 \mathrm{Ta}^{2/3}$.}
\label{fig:1}
\end{figure}

The result is an SDP which yields the desired optimal bound, but it is not
completely straightforward to compare the optimal
coefficients $\lambda_n$ to the optimal $\varphi(z)$
or $\tau(z)$ one might obtain from an analytical calculation. The reson for this is 
that the optimization problem includes constraints containing the integral of two
functions $f(z)$ and $g(z)$, one of which (for example $f$) is replaced in the course 
of the discretization by a sum of Dirac $\delta$-functions as 
$\sum_{n=1}^N f_n \delta(z-z_n)$ where the collocation points $z_n$ are given by
\begin{equation}
z_n = \frac{1}{2} \left[ 1+\cos \left( \pi \frac{n-1}{N-1} \right) \right]
~~~,~~~ n=1 ... N  
\label{eq:colloc}
.
\end{equation}
The integral $\int f(z) g(z) dz$ then becomes the sum $\sum_{n=1}^N f_n g(z_n)$. There
is no immediate relationship between $f(z)$ and the $f_n$ if the $f_n$ are chosen
such that the sum $\sum_{n=1}^N f_n g(z_n)$ approximates as closely as possible
the integral $\int f(z) g(z) dz$. For instance, these $f_n$ depend on the resolution 
$N$ and the collocation points $z_n$. The SDP as implemented returns a set of coefficients 
$a_n$ which appear in a representation of the first derivative of $\varphi(z)$
as $\varphi'(z) = \sum_{n=1}^N a_n \delta(z-z_n)$. The magnitude of these $a_n$ is not
directly connected to optimal background fields obtained from analytical calculations,
but the general dependence on space must be comparable.

Analytical treatment of infinite Prandtl number convection with the background
field method \cite{Doerin01,Consta99,Consta99b,Plasti05} used functions
$\tau(z)$ or $\varphi(z)$ whose first derivative is either constant or
piecewise constant, with a certain value for this derivative
in the central region of the layer and another value in two boundary layers
adjacent to the top and bottom boundaries. More sophisticated background fields
were also used in the non rotating problem \cite{Doerin06,Otto11}.
The thickness of the boundary layers
is imposed by the Rayleigh number. Fig. \ref{fig:1} shows for a few examples
the $a_n$, each $a_n$ plotted at the location $z_n$. The $a_n$ are approximately
constant for the $z_n$ near the center of the layer, but there is a rapid
variation of the $a_n$ within apparent boundary layers. More than one extremum
appears for no slip boundaries at large $\mathrm{Ta}$. 
The thickness of the boundary layer for no slip boundaries, defined from the
location of the minimal $a_n$, is compatible with the scaling in $\mathrm{Ta}^{-1/4}$ known from Ekman layers,
whereas the layers remain thicker for free slip boundary conditions. The demand
on resolution is therefore higher for no slip boundaries.

\newpage
\clearpage

\begin{figure}
\includegraphics[width=10cm]{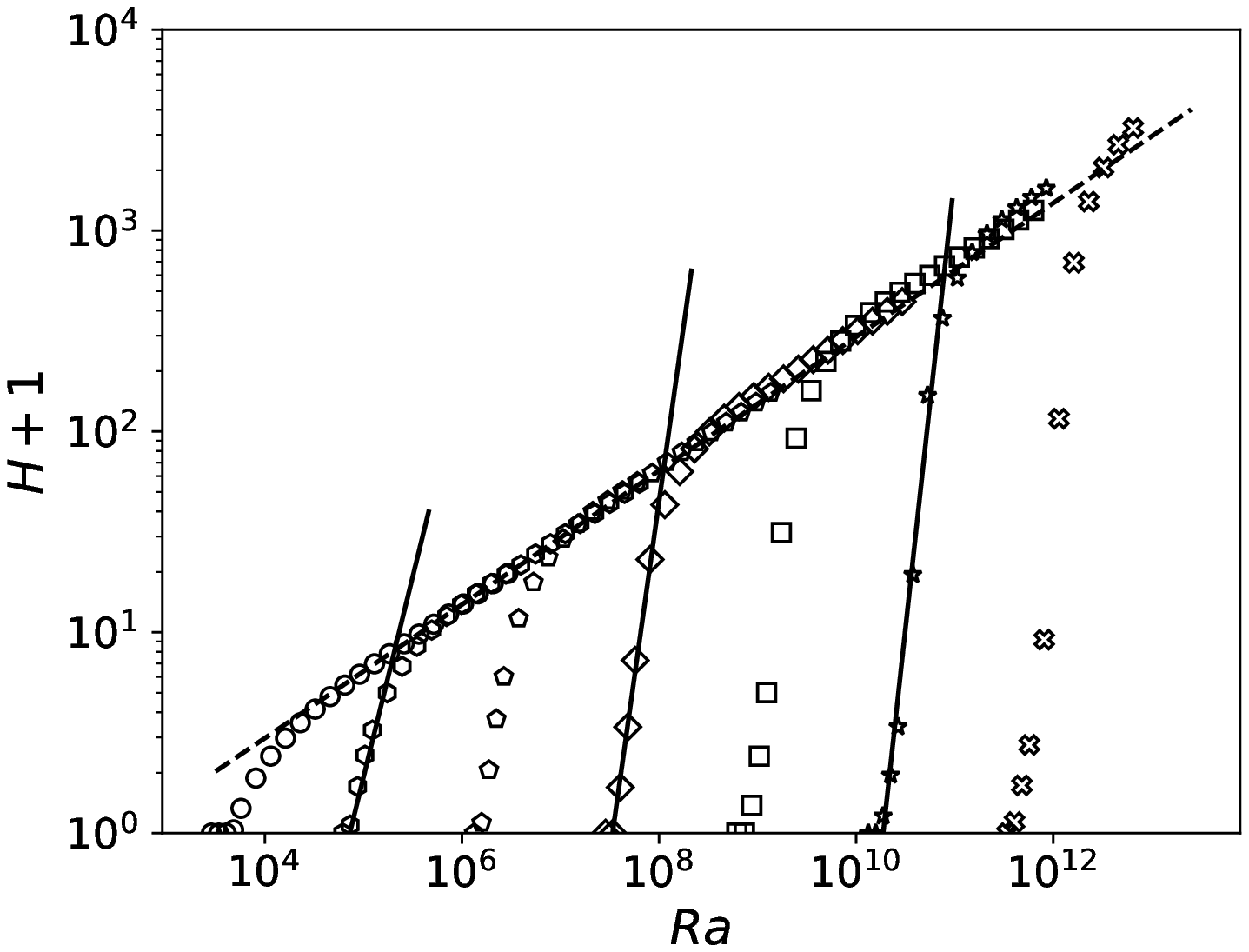}

\includegraphics[width=10cm]{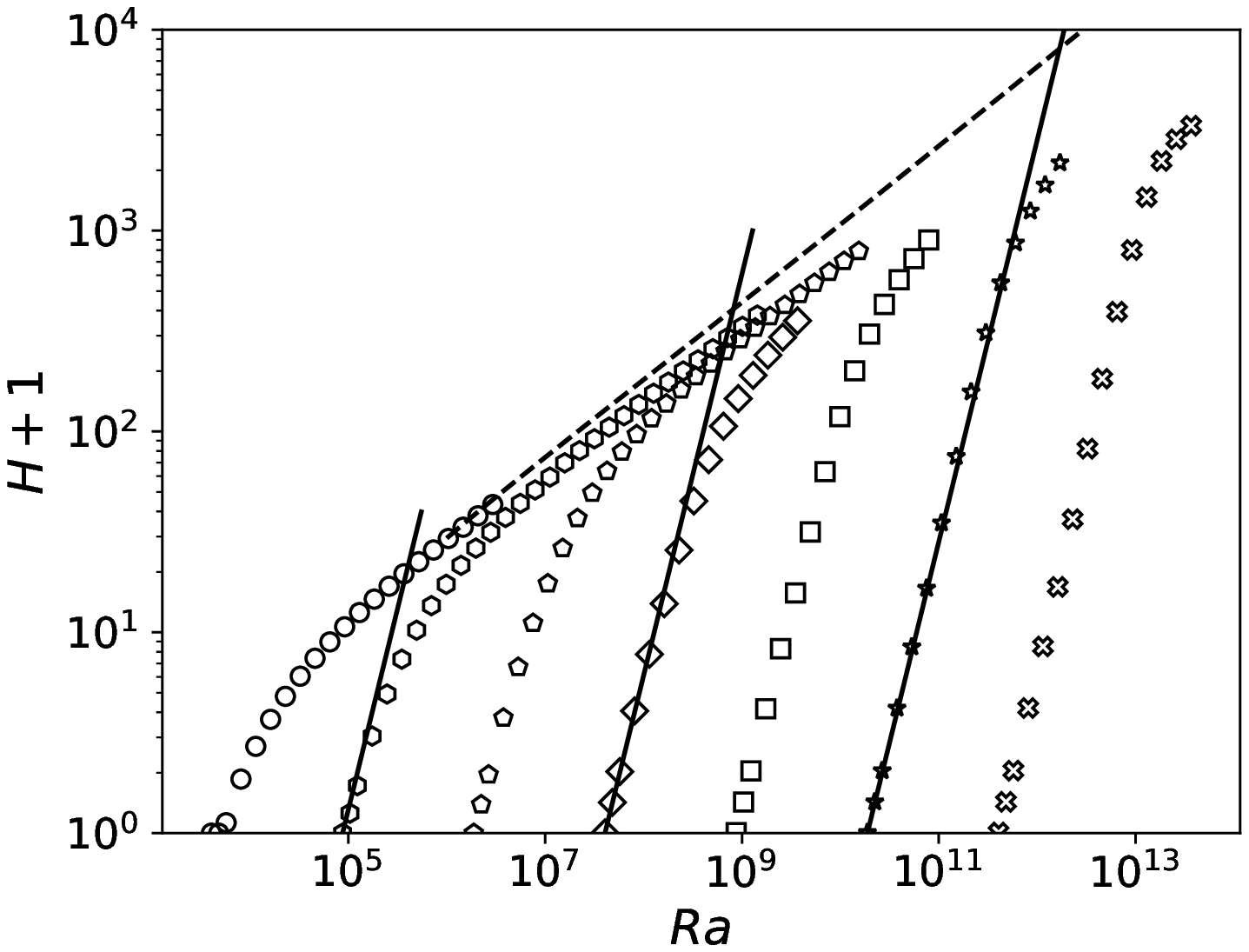}
\caption{
The bound on the Nusselt number $H+1$ as a function of $\mathrm{Ra}$ for
$\mathrm{Ta}=10^4$ (circles), $10^6$ (hexagons), $10^8$ (pentagons), $10^{10}$
(diamonds), $10^{12}$ (squares), $10^{14}$ (stars) and $10^{16}$ (crosses). The
upper panel is for no slip boundaries and the lower panel for stress free
boundaries. The dashed lines are fits to bounds obtained for $\mathrm{Ta}=0$
(see Refs. \onlinecite{Ierley06,Tilgne17b}) given by $H+1=0.101 \times \mathrm{Ra}^{0.4} + 0.70965 \times \mathrm{Ra}^{0.2} -
7.166$ for stress free boundaries and
$H+1=0.139 \times \mathrm{Ra}^{1/3}$ for no slip boundaries.
The continuous lines show functions
of the form $(\mathrm{Ra}/\mathrm{Ra}_\mathrm{crit})^\beta$ with
$\mathrm{Ra}_\mathrm{crit}$ computed from the asymptotically valid expression 
\cite{Chandr61} $\mathrm{Ra}_\mathrm{crit} = 8.73 \times \mathrm{Ta}^{2/3}$ for
stress free boundary conditions (upper panel) and obtained as part of the fitting
procedure
for no slip boundary conditions (lower panel). The exponent $\beta$ in the case
of no slip boundaries is 3 ($\mathrm{Ta}=10^6$), 3.5 ($\mathrm{Ta}=10^{10}$) and 
4.5 ($\mathrm{Ta}=10^{14}$), whereas $\beta=2$ for all lines drawn for stress
free boundaries.}
\label{fig:2}
\end{figure}

\newpage
\clearpage

\section{Results}
\label{section:Results}

The bound $H$ on the advective heat transport is obtained from the selection 
$Z = \langle v_z \theta \rangle$. The results of experiments or numerical
simulations are usually reported in terms of the Nusselt number, which is 
identical to $\langle \overline{v_z \theta} \rangle + 1$. For easier comparison,
fig. \ref{fig:2} shows $H+1$ as a function of $\mathrm{Ra}$ for various $\mathrm{Ta}$.
The graphs have the general appearance
familiar from experiments or numerical simulations: the bound $H$ is exactly zero
for $\mathrm{Ra}$ less than the critical Rayleigh number
$\mathrm{Ra}_\mathrm{crit}$ computed from linear stability analysis and the heat
transport is reduced by rotation compared with the non rotating flows at
Rayleigh numbers just above onset with a rapid increase of $H+1$ as a function of
$\mathrm{Ra}$, whereas at sufficiently large $\mathrm{Ra}$, $H+1$ increases more
slowly as a function of $\mathrm{Ra}$ at a rate asymptotically identical with
the non rotating case. For no slip boundaries, there is an overshoot in the sense that
the bound for rotating convection
is actually larger than that for the non rotating case in an intermediate
interval of $\mathrm{Ra}$. This
echoes the behaviour known from experiments and simulations in which the Nusselt
number of rotating convection may exceed the Nusselt number of non rotating
convection \cite{Zhong93,Zhong09,Schmit10}, a behaviour that is not observed
for stress free boundaries \cite{Schmit09}.

There does not seem to be any numerical data available for strictly infinite
$\mathrm{Pr}$, but the results in Ref. \onlinecite{King09} suggest that there is
no significant difference in the Nusselt number in going from $\mathrm{Pr}=7$ to
$\mathrm{Pr}=100$ so that it is meaningful to compare data obtained for
$\mathrm{Pr}$ equal to 7 or larger to the bounds for infinite $\mathrm{Pr}$. The
bounds for no slip boundaries in fig. \ref{fig:2} may then directly be compared
to the data compiled in figure 1 of Ref. \onlinecite{Julien16}. 
It is frequently attempted to fit power laws to experimentally or numerically
determined values of the Nusselt number.
Power laws are not a terribly good fit to either $H$ or the bounds $H+1$
in fig. \ref{fig:2} in the rotation
dominated regime, but if one insists on fitting power laws to the dependence of
$H+1$ on $\mathrm{Ra}$, one finds for no slip boundaries exponents of 2, 3.5 and
4.5 at approximately the Taylor numbers at which the exponents for the actual
Nusselt number at $\mathrm{Pr}=7$ are 1.2, 3 and 3.6 according to Ref. \onlinecite{Julien16}. The
bounds parallel the behaviour of the Nusselt number in that larger exponents are
necessary at larger $\mathrm{Ta}$, but the exponents obtained from fits to the
bounds are systematically larger than those obtained from the experiments.

For free slip boundaries, the dependence of $H+1$ on $\mathrm{Ra}$ for
$\mathrm{Ra}$ just above $\mathrm{Ra}_\mathrm{crit}$ is reasonably fitted by
$H+1 = (\mathrm{Ra}/\mathrm{Ra}_\mathrm{crit})^2$ at all $\mathrm{Ta}$. From
simulations \cite{Schmit10} at $10^2 \lesssim \mathrm{Ta} \lesssim 10^8$ one finds 
for $\mathrm{Pr}=7$ that the Nusselt number
just above onset behaves approximately as 
$(\mathrm{Ra}/\mathrm{Ra}_\mathrm{crit})^{6/5}$, so that the exponents fitting
the bounds again appear to be larger than what they should be if the bounds were
sharp.

It is also seen in figs. \ref{fig:1} and \ref{fig:2} that for stress free
boundaries, the non rotating behaviour is attained only at very large Rayleigh
numbers, if at all. For infinite $\mathrm{Pr}$, it is not obvious why the
behaviour of rotating and non rotating convection should approach each other at
any $\mathrm{Ra}$. We currently do not have a solid understanding of the
mechanism which determines the transition from convection dominated by Coriolis
force to convection nearly independent of rotation as $\mathrm{Ra}$ is
increased. \citet{King09} suggested that the transition occurs when the Ekman
layer and thermal boundary layer are equally thick. However, this criterion
cannot be generally applicable because the same transition is observed for
stress free and for no slip boundary conditions \cite{Schmit09,Schmit10} and
there are no Ekman layers near the stress free boundaries. The most intuitive
criterion is based on the Rossby number. If this number is large, the Coriolis
term is small compared with the advection term in the momentum equation.
\citet{Horn15} classified states of rotating convection with the help of the
Rossby number based on the free fall velocity,
$(g \alpha \Delta T h )^{1/2} / (\Omega h) \propto
(\mathrm{Ra}/(\mathrm{Pr}\mathrm{Ta}))^{1/2}$.
This number is always zero in the model of convection at infinite Prandtl
number, which is not surprising because there is no advection term left in the
momentum equation compared to which the Coriolis term could be small. 
It is also known \cite{Schmit09,Schmit10}
that if the Rossby number is based on the actual flow
velocity rather than the free fall velocity, the Rossby number does not allow
one to distinguish at $\mathrm{Pr}=0.7$ or $\mathrm{Pr}=7$
between convection dominated by
rotation from convection nearly independent of rotation. The balance between
Coriolis and rotation terms does not seem to be essential for the transition.

\begin{figure}
\includegraphics[width=8cm]{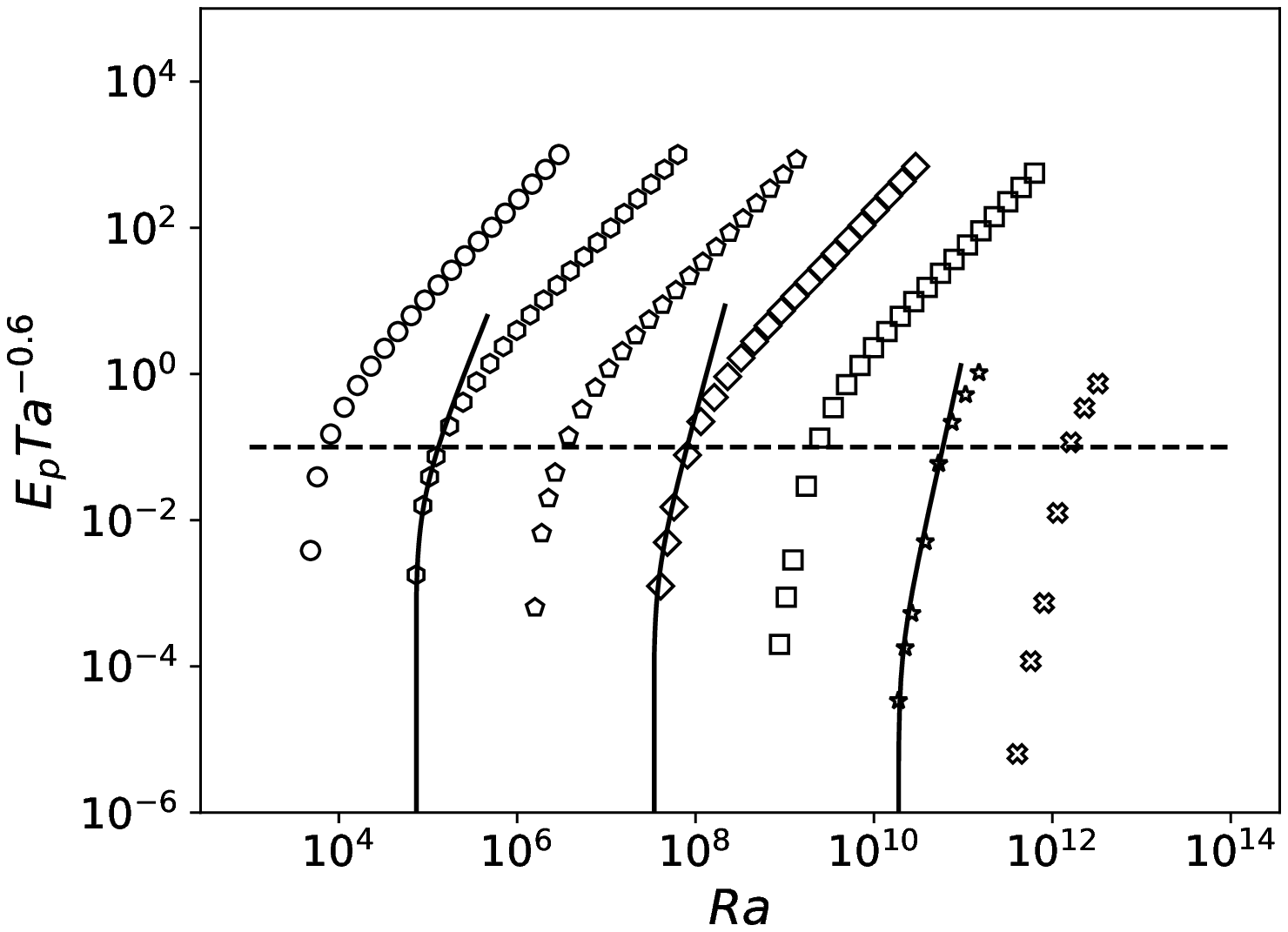}
\includegraphics[width=8cm]{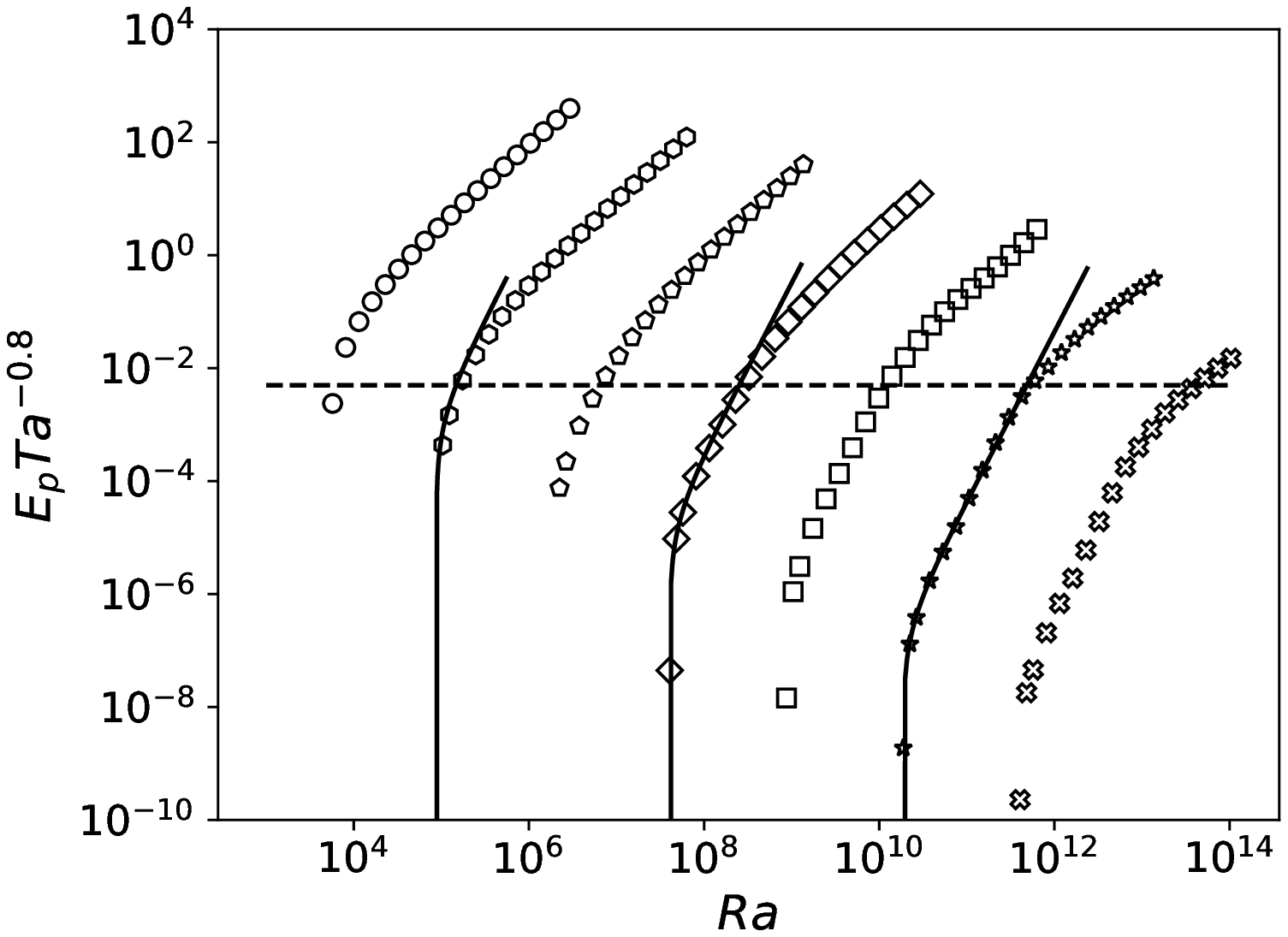}
\caption{
The bound $E_p$ on the poloidal energy shown in the combination
$E_p \mathrm{Ta}^{-0.6}$ for no slip boundaries (left panel) and
$E_p \mathrm{Ta}^{-0.8}$ for stress free boundaries (right panel) with the same
symbols as in fig. \ref{fig:2}. The continuous lines indicate the functions
$(\mathrm{Ra}/\mathrm{Ra}_\mathrm{crit})
[(\mathrm{Ra}/\mathrm{Ra}_\mathrm{crit})^\beta -1] \mathrm{Ta}^{1/3}$
for $\beta=2$, $3.5$ and $4.5$ in the left panel and $\beta=2$ in the right
panel. The dashed horizontal lines are located at
$E_p \mathrm{Ta}^{-0.6}=0.1$ (left panel) and
$E_p \mathrm{Ta}^{-0.8}=5 \times 10^{-3}$ (right panel).}
\label{fig:3}
\end{figure}

Another reasoning, put forward by \citet{Schmit09}, combines the asymptotic
behaviour of the Nusselt number at small and large $\mathrm{Ra}$ for fixed
$\mathrm{Ta}$ to predict that the transition should occur if
\begin{equation}
(E_\mathrm{pol} + E_\mathrm{tor}) \mathrm{Ta}^{-1} = \mathrm{const.}
\label{eq:criterion}
\end{equation}
This criterion, which is valid for both types of boundary conditions
\cite{Schmit10}, is independent of $\mathrm{Pr}$ and can be immediately extended
to infinite $\mathrm{Pr}$. This criterion also motivates the form in which the bounds
are presented in the following two figures.

\begin{figure}
\includegraphics[width=8cm]{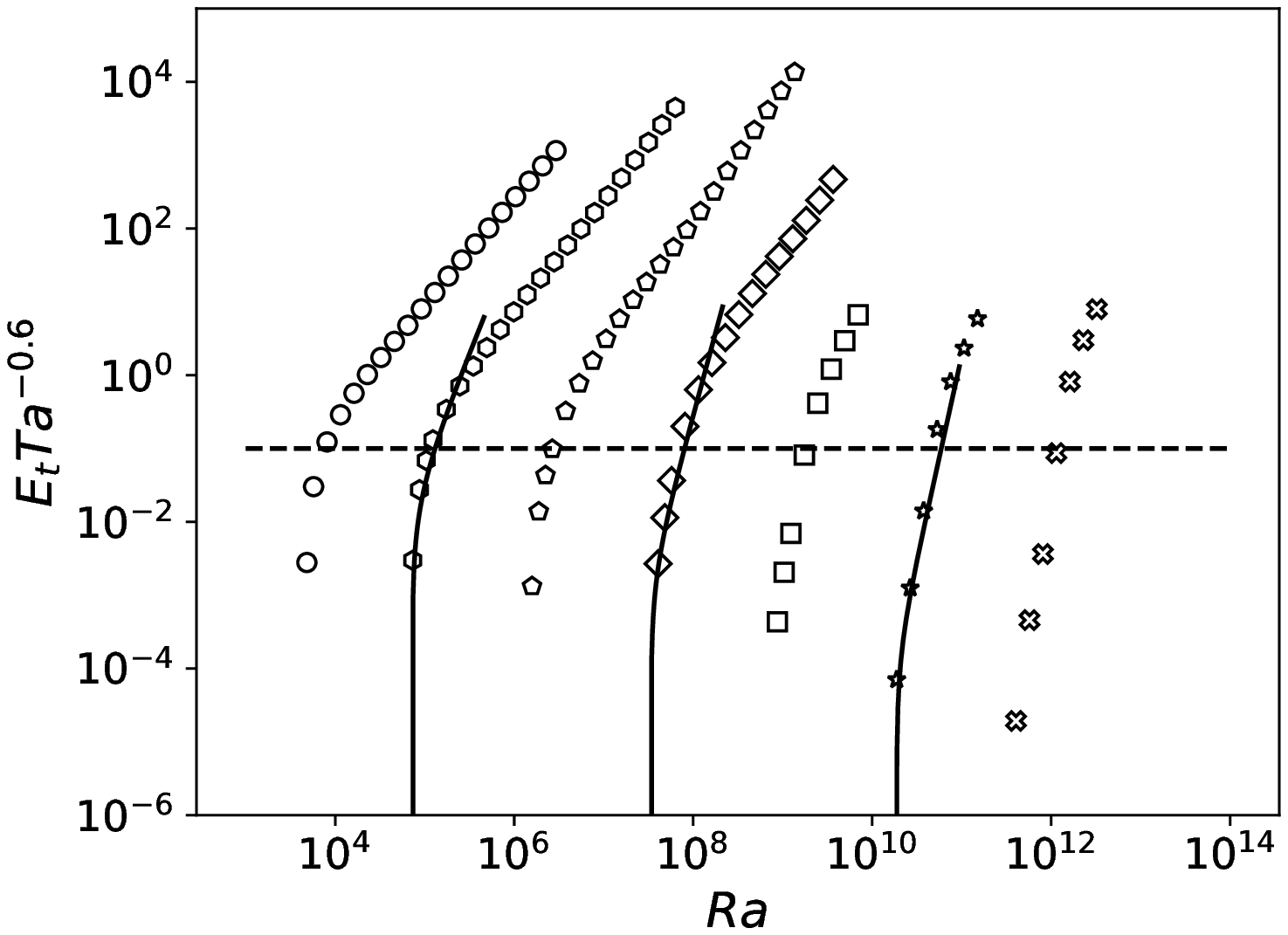}
\includegraphics[width=8cm]{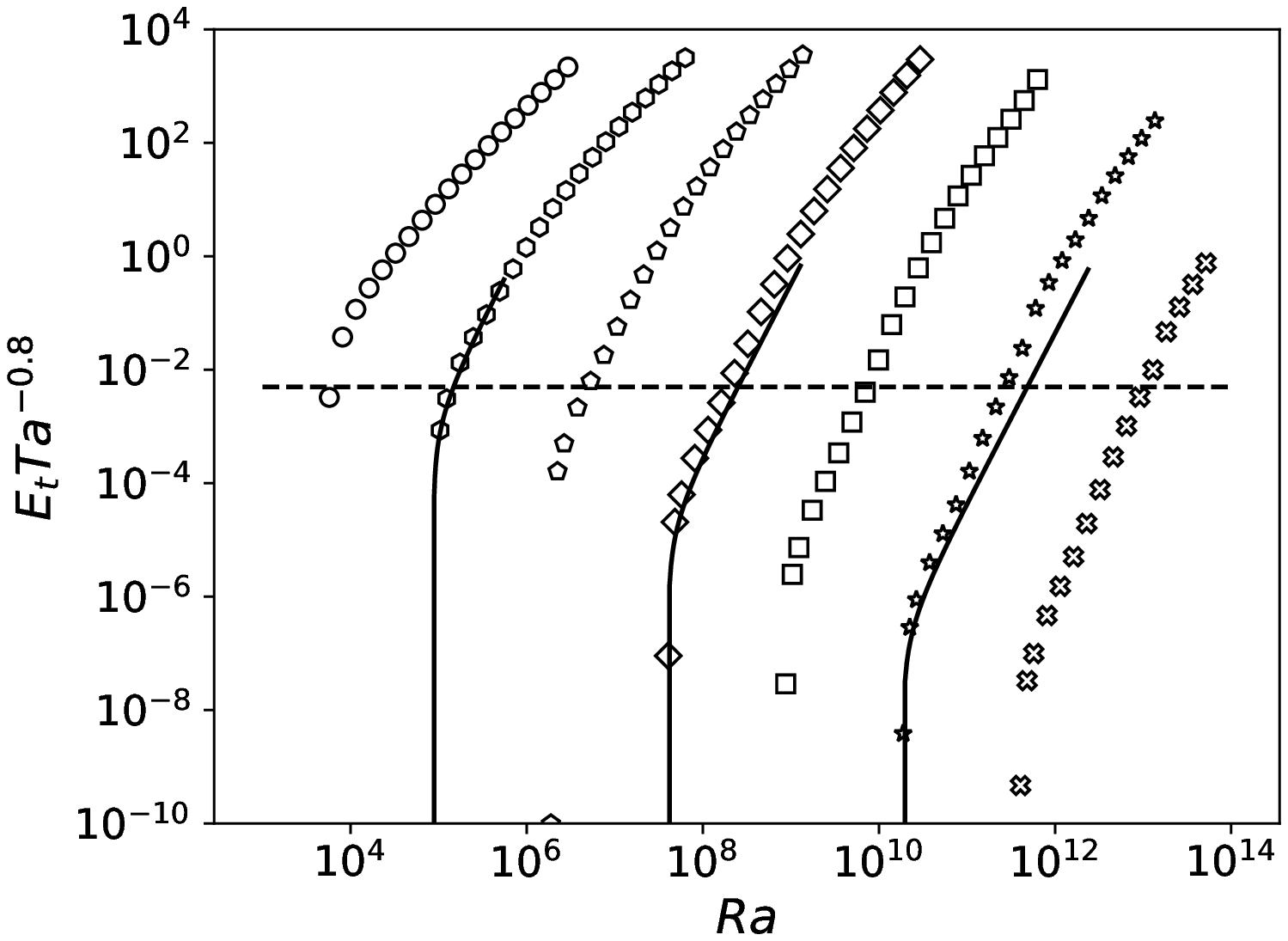}
\caption{
The bound on the toroidal energy, $E_t$, in the same representation as $E_p$ in
fig. \ref{fig:3}, with the same symbols and the same dashed and continuous
lines. The left panel is for no slip boundaries and the right panel for stress
free boundaries.}
\label{fig:4}
\end{figure}

Figs. \ref{fig:3} and \ref{fig:4} show the bounds on poloidal and toroidal
energies, $E_p$ and $E_t$, obtained from
$Z= \langle \frac{1}{2} |\nabla \times \nabla \times (\phi \bm{\hat z})|^2 \rangle$ and
$Z= \langle \frac{1}{2} |\nabla \times (\psi \bm{\hat z})|^2 \rangle$, respectively.
At any $\mathrm{Ta}$, these bounds show again two regions of $\mathrm{Ra}$ in
which rotation is either suppressing convection or not. To better identify the interval of
$\mathrm{Ra}$ dominated by rotation, it helps to make a connection with the
bounds on the heat transport. The $z$ component of velocity is zero at the
boundaries so that the bound $H$ implies by virtue of the Poincar\'e inequality
a bound on the poloidal energy \cite{Tilgne17b} as
\begin{equation}
E_\mathrm{pol} \leq \frac{1}{2 \pi^2} \mathrm{Ra} H 
.
\label{eq:ineq_Ekin}
\end{equation}
However, one expects a stricter bound to be valid. Taking the dot product of the
momentum equation (\ref{eq:NS}) with $\bm v$ and integrating over space leads to
\begin{equation}
\sum_{ij} \langle (\partial_j v_i)^2 \rangle = \mathrm{Ra} \langle v_z \theta \rangle
.
\label{eq:E_budget}
\end{equation}
If $\lambda_c$ is a characteristic length scale of the flow, one expects
$\sum_{ij} \langle (\partial_j v_i)^2 \rangle \sim (E_\mathrm{pol} +
E_\mathrm{tor}) / \lambda_c^2$
and therefore
$E_\mathrm{pol} \leq \mathrm{Ra} H / \lambda_c^2$. 
At $\mathrm{Ra}$ just above critical, the typical length scale of rotation is
the wavelength of the most unstable mode in linear stability analysis, which
leads to\cite{Chandr61} $\lambda_c^2 \propto \mathrm{Ta}^{-1/3}$. Replacing $H$ by its power
law fit deduced from fig. \ref{fig:2} and using
$\mathrm{Ra}/\mathrm{Ra}_\mathrm{crit}$ as variable, one expects the existence
of a bound $E_p$ of the form
\begin{equation}
E_\mathrm{pol} \leq E_p \propto \frac{\mathrm{Ra}}{\mathrm{Ra}_\mathrm{crit}}
\left[ \left( \frac{\mathrm{Ra}}{\mathrm{Ra}_\mathrm{crit}} \right)^\beta -1
\right] \mathrm{Ta}^{1/3}
\label{eq:Epol_fit}
\end{equation}
This functional form, with a prefactor equal to 1, fits well the dependence of
$E_p$ on $\mathrm{Ra}$ for rotating convection with
both types of boundary conditions and $\mathrm{Ra}$ close to $\mathrm{Ra}_\mathrm{crit}$.
The bound $E_p$ obtained from the SDP is
therefore sensitive to the variation of the characteristic length scale of
convection with $\mathrm{Ta}$, in agreement with analytically computed bounds
\cite{Tilgne22}.

Eq. (\ref{eq:Epol_fit}) fails to be a good fit if $\mathrm{Ra}$ is large enough,
which marks the upper edge of the interval of $\mathrm{Ra}$ in which the
bound $E_p$ is controlled by the Coriolis force. 
Fig. \ref{fig:3} shows that the bound on $E_\mathrm{pol}$ is indicative of
rapidly rotating convection only if $E_p \mathrm{Ta}^{-0.6} \leq 0.1$ for no slip and
$E_p \mathrm{Ta}^{-0.8} \leq 5 \times 10^{-3}$ for stress free boundaries. 
Fig. \ref{fig:4} repeats all the previous considerations for the toroidal
energy. For no slip boundaries, the same function as in eq. (\ref{eq:Epol_fit})
approximates also the bounds for the toroidal energy in the rapidly rotating
regime, which implies near equipartition between the poloidal and toroidal energies in as far as bounds
are concerned. In the case of stress free boundaries, the bound on toroidal
energy $E_t$ exceeds the bound on poloidal energy. For both boundary conditions, $E_t$
exceeds $E_p$ in the high Rayleigh number regime even though the toroidal energy is
zero in the non rotating case. At finite $\mathrm{Pr}$, simulations \cite{Horn15}
show that the fraction of toroidal to total energy decreases with $\mathrm{Ra}$
at large $\mathrm{Ra}$. This behavior is not reproduced by the bound $E_t$, possibly
because eq. (\ref{eq:psi}) prevents the fraction of toroidal to total energy from
decreasing to zero which would mean that in this particular respect, rotating convection
at infinite and finite $\mathrm{Pr}$ are fundamentally different from each other. 
Fig. \ref{fig:4} also shows that for no slip boundaries, the condition
$E_t \mathrm{Ta}^{-0.6} \leq \mathrm{const.}$ reasonably delimits the region in parameter
space in which the
bound $E_t$ is determined by the length scale of rapidly rotating convection near
onset, $\lambda_c$. Combined with the bounds for poloidal energy, this leads to
$(E_p +E_t) \mathrm{Ta}^{-0.6} \leq \mathrm{const.}$ as a criterion for the bounds
being mainly controlled by the Coriolis term, which contains a different exponent than
eq. (\ref{eq:criterion}).

\section{Summary}

Predictions of heat transport in turbulent convection necessarily rely on
empirically motivated assumptions. An early approach assumed that the thermal
boundary layers are marginally stable which leads in the non rotating case to
\cite{Malkus54,Howard66} $\mathrm{Nu} \propto \mathrm{Ra}^{1/3}$. It is possible to derive
bounds that come at least to within logarithmic factors of $\mathrm{Ra}^{1/3}$
for convection at infinite Prandtl number
\cite{Chan71,Ierley06,Consta99,Doerin06,Otto11}.
Bounds for convection at arbitrary Prandtl number \cite{Doerin96} are no better
than $\mathrm{Nu} \propto \mathrm{Ra}^{1/2}$. It is tempting to 
ascribe this difference to the different nature of the equations of evolution:
the momentum equation becomes a diagnostic equation at infinite $\mathrm{Pr}$,
effectively reducing the problem to a single equation for the temperature field.
It is then interesting to investigate whether similarly tight bounds can be
derived for rotating convection at infinite $\mathrm{Pr}$.

Attempts at analytically deriving bounds on heat transport in rotating
convection at infinite $\mathrm{Pr}$ have been based on the background field
method \cite{Doerin01,Consta99b}
which enforces constraints derived from the surface average over
horizontal planes of the temperature equation and the volume average of the
temperature equation multiplied by $\theta$. The numerical search of an optimal
bound satisfying constraints of
this form leads to an SDP. For $\mathrm{Ta}=0$ and no slip boundaries, the SDP
technique finds an upper bound for the heat transport which varies as a function
of $\mathrm{Ra}$ as $\mathrm{Ra}^{1/3}$, at least to within logarithmic factors
\cite{Nobili17}. Heat transport is reduced by rotation.
This reduction is reproduced in the bound computation, but not by as much as is
observed in numerical computations at finite but large $\mathrm{Pr}$. The SDP
finds the correct $\mathrm{Ra}$ for the onset of convection, but at larger
$\mathrm{Ra}$, the bound on $\mathrm{Nu}$ increases more rapidly with
$\mathrm{Ra}$ than the $\mathrm{Nu}$ obtained from simulations. The mere fact
that momentum equation is reduced to a diagnostic equation is therefore no
guarantee that the background method will provide us with satisfactory bounds.

The bounds obtained from the SDP are nonetheless stricter than the bounds
obtained from analytical work in as far as the dependence on rotation is
concerned. Both types of calculations enforce the same constraints, but the SDP
optimizes the background field profile and does not rely on simplifying
estimates to ensure the validity of inequalities analogous to (\ref{eq:p7Mitte})
usually called spectral constraint in the context of the background method.
Linear stability theory computes a critical Rayleigh number 
$\mathrm{Ra}_\mathrm{crit}$ which asymptotically scales with $\mathrm{Ta}$ as
$\mathrm{Ra}_\mathrm{crit} \propto \mathrm{Ta}^{2/3}$ at large $\mathrm{Ta}$.
For stress free boundaries, the result from the SDP may be presented as 
$\mathrm{Nu} \lesssim \left( \mathrm{Ra} / \mathrm{Ra}_\mathrm{crit} \right)^2$
ignoring numerical prefactors. The rotation dependent bounds obtained
analytically are either 
$\mathrm{Nu}-1 \lesssim \mathrm{Ra}^{1/2} \left( \mathrm{Ra} /
\mathrm{Ra}_\mathrm{crit} \right)^{1/2}$
from Ref. \onlinecite{Tilgne22} or
$\mathrm{Nu}-1 \lesssim \mathrm{Ra}^{1/2} \left( \mathrm{Ra} /
\mathrm{Ra}_\mathrm{crit} \right)^{3/2}$
from Ref. \onlinecite{Consta99b} in the rotation dominated regime. 
Both of these bounds impose a weaker reduction
of $\mathrm{Nu}$ with $\mathrm{Ta}$ than the bound from the SDP. The
expressions for no slip boundaries are more complicated since the analytical
result \cite{Consta99b} for the bound on $\mathrm{Nu}-1$ is a sum of two terms,
one varying as 
$\mathrm{Ra}^{1/2} \left( \mathrm{Ra} / \mathrm{Ra}_\mathrm{crit}
\right)^{3/2}$, the other as
$\mathrm{Ra}^{5/4} \left( \mathrm{Ra} / \mathrm{Ra}_\mathrm{crit}
\right)^{3/4}$. Both terms decrease more slowly with increasing $\mathrm{Ta}$
as the $\left( \mathrm{Ra} / \mathrm{Ra}_\mathrm{crit} \right)^\beta$ with
$\beta > 2$ in fig. \ref{fig:2}. 

The SDP bounds the advective heat flux to zero for 
$\mathrm{Ra} < \mathrm{Ra}_\mathrm{crit}$. The SDP bounds also reflect
several qualitative effects known from experiments and numerical simulations,
such as a characteristic length scale varying as $\mathrm{Ta}^{-1/6}$ in the
rotation dominated regime. In the case of no slip boundaries, the heat
transport in rapidly rotating convection and in some interval of Rayleigh numbers
exceeds the heat transport of non rotating convection at the same Rayleigh
numbers. The exponent $\beta$ in 
$\mathrm{Nu} \lesssim \left( \mathrm{Ra} / \mathrm{Ra}_\mathrm{crit} \right)^\beta$
in the bounds for no slip boundary conditions increases with $\mathrm{Ta}$ 
(see fig. \ref{fig:2}), just
as it does in fits to experimentally and numerically determined
$\mathrm{Nu}$.

Rotating convection at infinite Prandtl number thus is a system in which
significant improvement over existing analytical results is possible within the
background method. It is also a system which is a promising testing ground for
bounding methods involving additional constraints. It is possible to formulate
an SDP which takes into account more constraints \cite{Cherny14}, but at the
cost of a significantly larger computational burden. A method of this type was
recently demonstrated in the context of nonlinear stability analysis
\cite{Fuente22}. Rotating convection at infinite Prandtl number is a non trivial
fluid dynamic system, which is relatively simple in the sense that it is
governed by an equation for a single scalar field, and for which the known
bounds do not scale with the control parameters in the same way as the results
of numerical computations, so that this system promises to be particularly
rewarding for any improved bounding method.


%

\end{document}